\documentclass[aps,prb,times,twocolumn,a4paper]{revtex4-1}

\usepackage{colortbl,amsfonts,amsmath,amssymb}
\usepackage{hyperref,tikz,tabularx,hhline}
\usepackage{float,graphicx,dcolumn,stmaryrd}
\usepackage{subfigure,wasysym}

\usepackage{hyperref} 
\hypersetup{
colorlinks = True,
linkcolor = red,
urlcolor = blue,
citecolor = blue
}

\date{\today}

\begin{document}

\title{A Kagome Map of Spin Liquids:\\ from XXZ to Dzyaloshinskii-Moriya Ferromagnet}
\author{Karim Essafi}
\email{karim.essafi@oist.jp}
\author{Owen Benton}
\email{owen.benton@oist.jp}
\author{L.~D.~C. Jaubert}
\email{ludovic.jaubert@oist.jp}
\affiliation{Okinawa Institute of Science and Technology Graduate University, Onna-son, Okinawa 904-0495, Japan}

\begin{abstract}
The kagome lattice sits at the crossroad of present research efforts in quantum spin liquids, chiral phases, emergent skyrmion excitations and anomalous Hall effects to name but a few. In light of this diversity, our goal in this paper is to build a unifying picture of the underlying magnetic degrees-of-freedom on kagome. Motivated by a growing mosaic of materials, we especially consider a broad range of nearest-neighbour interactions consisting of Dzyaloshinskii-Moriya as well as anisotropic ferro$-$ and antiferromagnetic coupling. We present a three-fold mapping on the kagome lattice which transforms the celebrated Heisenberg antiferromagnet and XXZ model onto two lines of time-reversal invariant Hamiltonians. The mapping is exact for classical and quantum spins alike, \textit{i.e.} it preserves the energy spectrum of the original Heisenberg and XXZ models. As a consequence, at the classical level, all phases have an extensive ground-state degeneracy. These ground states support a variety of phenomena such as ferromagnetically induced pinch points in the structure factor and the possibility for spontaneous scalar chirality. For quantum spin$-1/2$, the XXZ model has been recently shown to be a quantum spin liquid. Applying our three-fold mapping to the XXZ model gives rise to a connected network of quantum spin liquids, centered around a paragon of quantum disorder, namely the Ising antiferromagnet. We show that this quantum disorder spreads over an extended region of the phase diagram at linear order in spin wave theory, which overlaps with the parameter region of Herbertsmithite ZnCu$_3$(OH)$_6$Cl$_2$. We conclude this work by discussing the connection of our results to the chiral spin liquids found on kagome with further nearest-neighbour interactions, and to the recently synthesized ternary intermetallic materials.
\end{abstract}


\maketitle


Competing interactions have proven able to stabilize extended phases where chirality could be encoded in the spin texture, \textit{i.e.} coming from the collective behaviour of spins. This spin-chirality is responsible for phenomena as varied as the anomalous Hall effect~\cite{Ohgushi00a,Yoshii00a,Taguchi01a,Machida10a}, multiferroicity~\cite{Katsura05a} and possibly high-$T_{c}$ superconductivity~\cite{Wen89a}. In this context, kagome systems are fertile soil for exotic spin textures. Motivated by a growing number of materials~\cite{Hiroi01a,Shores05a,Colman08a,Okamoto09a,Aidoudi11a,Han14a,Gorbunov14a,Nakamura15a}, the kagome lattice, whose name comes from a traditional Japanese woven bamboo pattern~\cite{Syozi51a}, has attracted the attention of chemists, experimentalists and theorists alike. The classical kagome antiferromagnet is a canonical example of order-by-disorder~\cite{Chalker92a}, a counter-intuitive mechanism where order is induced by fluctuations~\cite{Villain80a}. As for its quantum counterpart, it is one of the few models that has been confirmed to be a quantum spin liquid by a gamut of complementary approaches~\cite{Hermele08a,Yan11a,Iqbal11a,Jiang12a,Depenbrock12a}. Recently, the kagome lattice has also been shown to support examples of the long-sought Kalmeyer-Laughlin chiral spin liquid~\cite{Bauer14a,Gong14a,He14a,Gong15a}, a bosonic analogue of the fractional quantum Hall effect with anyonic excitations~\cite{Kalmeyer87a,Kalmeyer89a}.\\

Our present work sits at the frontier of these ideas of unconventional phenomena, spin liquids and chiral phases. We unveil a three-fold mapping between kagome spin liquids, summarized in Figs.~\ref{fig:Map} and~\ref{fig:PD}, which is exact both at the classical and quantum level. This mapping brings into a general framework the well-known Heisenberg antiferromagnet and XXZ models, together with a continuously connected network of systems with Dzyaloshinskii-Moriya and anisotropic ferromagnetic couplings. All interactions are time-reversal ($\mathcal{T}$) invariant and between nearest neighbours (see Hamiltonian~(\ref{eq:ham})). For the end points of this connected network, $\mathcal{T}$ symmetry can be spontaneously broken in the classical ground state, giving rise to finite scalar chirality. The Heisenberg antiferromagnet maps onto a pair of systems characterized by \textit{ferromagnetic} pinch points in their structure factors, signatures of  algebraic correlations constrained by an effective local flux conservation. Interestingly for quantum spin$-1/2$, our work puts the Ising antiferromagnet at the centre of this connected network of quantum spin liquids, shedding a new light on the reluctance of this model to order~\cite{Moessner00b}. On the experimental front, our phase diagram includes the Herbertsmithite compound ZnCu$_3$(OH)$_6$Cl$_2$ which sits at the tip of an extended region of quantum disorder within the framework of linear spin wave theory. Our work is also motivated by the experimental possibility to explore a broad range of anisotropic interactions in the recently synthesized rare-earth kagome materials Dy$_{3}$Ru$_{4}$Al$_{12}$~\cite{Gorbunov14a} and Yb$_{3}$Ru$_{4}$Al$_{12}$~\cite{Nakamura15a} and in optical lattices~\cite{Jo12a,Struck13a}.\\


\section{Presentation}

\subsection{Model}

\begin{figure}[t!]
\begin{center}
\includegraphics[width=\columnwidth]{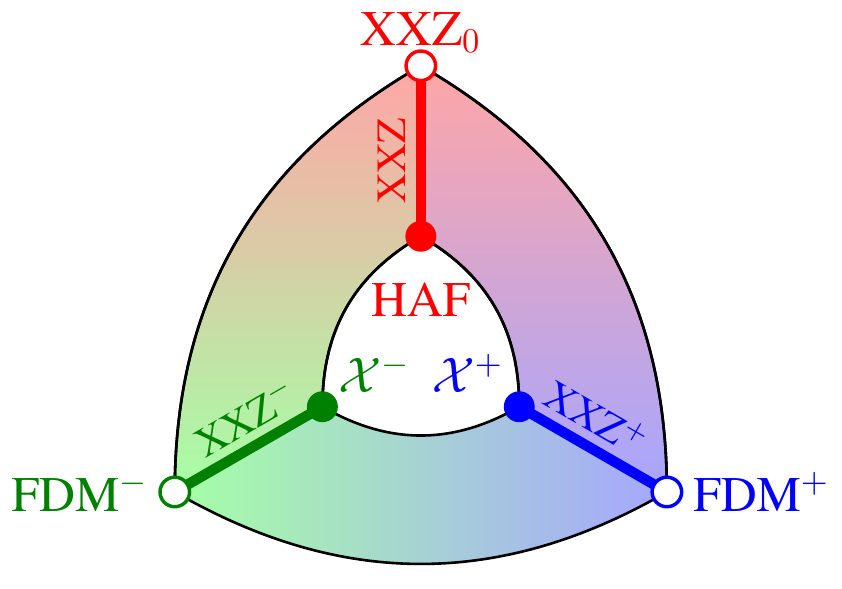}\\
\includegraphics[width=\columnwidth]{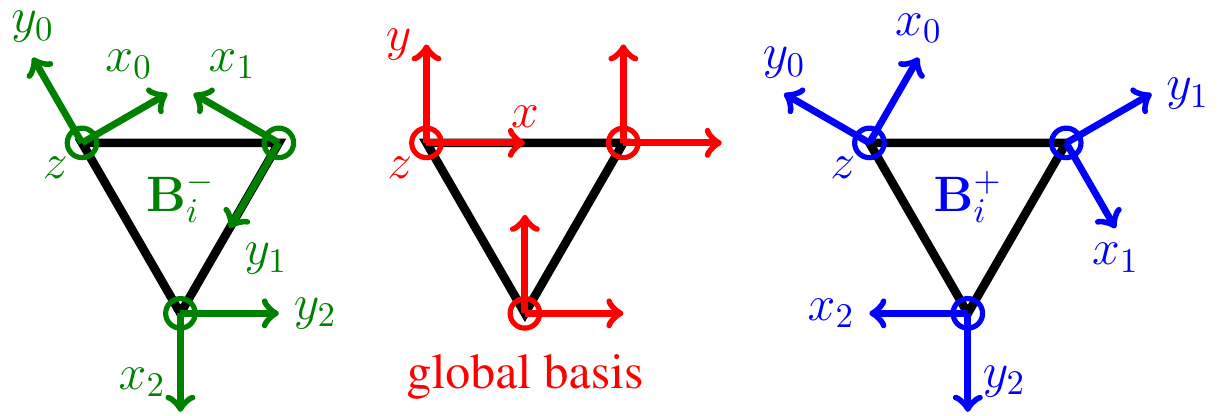}
\end{center}
\caption{\textbf{Three-fold mapping between kagome spin liquids} -- \textit{Top:} We show the existence of an exact one-to-one mapping, made of local proper spin rotations (bottom panels), between the celebrated Heisenberg antiferromagnet (HAF) and two novel spin liquids $\mathcal{X}^{\pm}$ with opposite vector chirality. By tuning the anisotropy coupling $\delta$ of Eqs.~(\ref{eq:XXZ}) and~(\ref{eq:XXZpm}), our mapping directly extends onto the anisotropic XXZ model. Its chiral counterparts (named XXZ$^{\pm}$) share the same extensive ground-state degeneracy as the XXZ model, until the end point $\delta=-1/2$ (FDM$^{\pm}$), which belongs to the ferromagnetic model with Dzyaloshinskii-Moriya interactions, and where chirality becomes scalar. \textit{Bottom:} The local bases are rotated by $\mp \frac{2\pi}{3}$ around the $z-$axis when moving from $\mathbf{B}_0^{\pm}\rightarrow\mathbf{B}_1^{\pm}\rightarrow\mathbf{B}_2^{\pm}$. The $z-$axis are the same for all bases, which are right-oriented. 
}
\label{fig:Map}
\end{figure}

We focus on the nearest-neighbour Hamiltonian with anisotropic XXZ and Dzyaloshinskii-Moriya interactions:
\begin{align}
\mathcal{H} =  \sum_{\langle ij \rangle} J_{\perp}\, \vec{S}_i^\perp \cdot \vec{S}_j^\perp + J_{z}\, S_i^z S_j^z + D\, \vec{z} \cdot \left(\vec{S}_i \times \vec{S}_j \right).
\label{eq:ham}
\end{align} 
We shall first consider classical Heisenberg spins of unit length $|\vec{S}_i|=1$ with in-plane components $\vec{S}_i^\perp = (S_i^x, S_i^y)$ for a system of $N$ spins. The sublattice indices and Cartesian bases are given in Fig.~\ref{fig:Map}. For perfect kagome symmetry, the Dzyaloshinskii-Moriya vector is restricted along the unit vector $\vec z$, orthogonal to the kagome plane~\cite{Moriya60a,Elhajal02a}, using the clockwise convention for choosing the  pairs of spins around the triangles.\\

\subsection{Heisenberg antiferromagnet (HAF)}

Parametrized by $J_{\perp}=J_{z}=J>0$ and $D=0$, the extensively degenerate ground-state manifold of the HAF is locally constrained by a magnetization flux conservation. This constraint appears clearly if the Hamiltonian is rewritten as
\begin{align}
\mathcal{H}_{\rm HAF} = J \sum_{\langle ij \rangle} \vec{S}_i \cdot \vec{S}_j = \frac{J}{2} \sum_{\Delta} \left( \sum_{i=0}^{2} \vec{S}_i \right)^2 \,-\,NJ,
\label{eq:HAF}
\end{align}
where the flux conservation takes the form of a null magnetization on all triangles $\Delta$: $\sum_{i=0}^{2} \vec{S}_i =0$.\\

\begin{figure*}[t!]
\subfigure[\large $\qquad \langle \vec S^{\perp}(\mathbf{q}) \cdot \vec S^{\perp}(-\mathbf{q})\rangle_{\rm HAF}$]{
\includegraphics[width=\columnwidth]{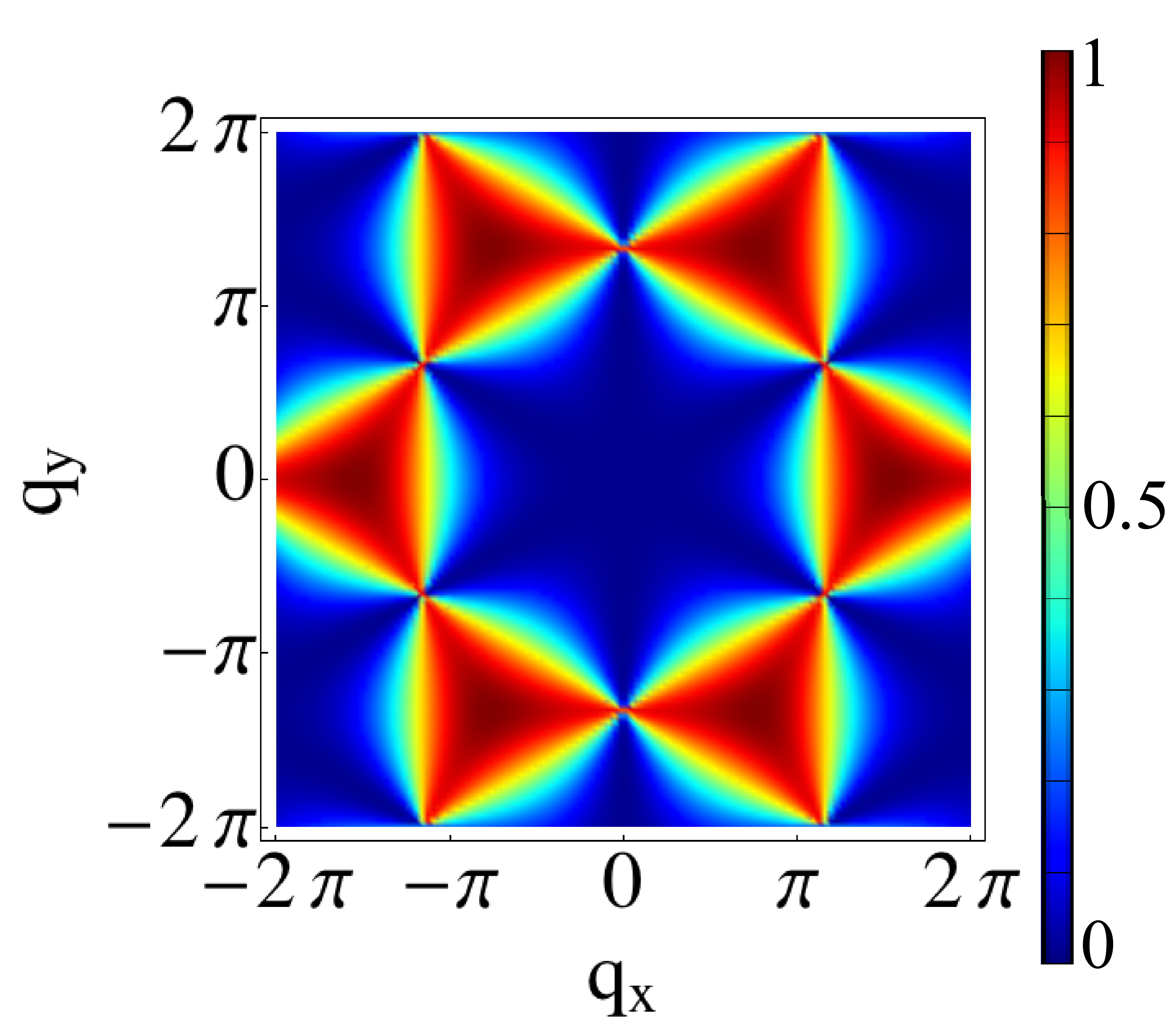}}
\subfigure[\large $\qquad \langle \vec S^{\perp}(\mathbf{q}) \cdot  \vec S^{\perp}(-\mathbf{q})\rangle_{\mathcal{X}^{\pm}}$]{
\includegraphics[width=\columnwidth]{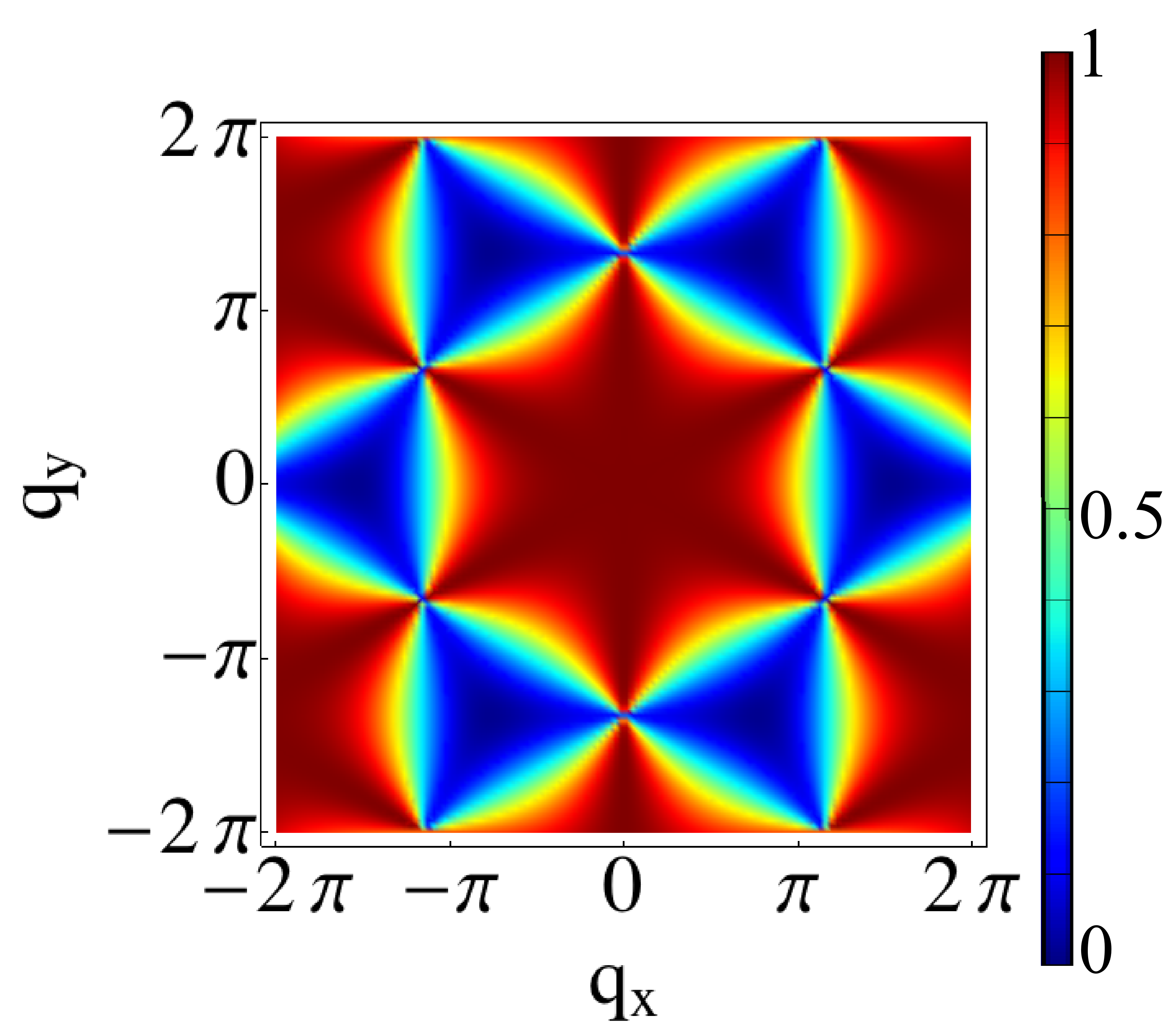}}
\caption{\textbf{Structure factor of the HAF (a) and $\mathcal{X}^{\pm}$ (b) spin liquids} -- The Fourier transforms of the spin correlations have been computed using the method developed by Henley~\cite{Henley05a} for Coulomb spin liquids in which the local constraints are enforced by a projection operator in reciprocal space. We have considered the planar spin components $\vec S^{\perp}$ where ``pinch point'' structures are formed in the centre of the Brillouin zones, characteristic of the local flux conservation. The structure factors clearly illustrate the underlying equivalence of the models, and the difference of their in-plane fluctuations; antiferromagnetic in the HAF and ferromagnetic in the $\mathcal{X}^{\pm}$ models. Only one figure is shown for the $\mathcal{X}^{\pm}$ phases because they cannot be distinguished by the structure factor of the planar spin components. The colour scales are fixed by the maximum of intensity on each figure.
}
\label{fig:Sq}
\end{figure*}

\section{Three-fold mapping}

The peculiarity of the HAF lies in the form of its Hamiltonian~(\ref{eq:HAF}). The idea of this paper is to find a one-to-one mapping (automorphism) of the spin degrees-of-freedom which gives a Hamiltonian that can be re-written in the same form, while conserving the kagome symmetry and the spin unit-length, without imposing any spurious constraints.

To ensure the spin unit length $|\vec S_{i}|=1$, we consider \textit{local} transformations $\Gamma$ acting on each spin independently, \textit{i.e.} transformations from the global basis to a local one, $\mathbf{B}_i$: $\vec S_{i}^{\,\mathbf{B}_i}= \Gamma^{\mathbf{B}_i}\vec S_{i}$ with $|\vec S_{i}^{\,\mathbf{B}_i}|=1$. Then for the transformation to be non-trivial -- \textit{i.e.} for $\mathbf{B}_i$ to be non-uniform -- and to respect translation invariance, we attach one basis $\mathbf{B}_i$ to each kagome sublattice. As a result, there are only two transformations respecting the space group symmetry of the kagome lattice. They are made of local proper rotations as illustrated in Fig.~\ref{fig:Map}. They transform the HAF into the following models which we name $\mathcal{X}^{\pm}$
\begin{align}
\mathcal{H}_{\mathcal{X}^{\pm}}
&= \frac{J}{2} \sum_{\Delta} \left( \sum\limits_{i=0}^{2} \vec{S}_i^{\,\mathbf{B}_i^{\pm}} \right)^2 \,-\,NJ
\label{eq:X1}\\
&= -\frac{J}{2} \sum_{\langle i,j\rangle} \left[ \vec{S}_i^\perp \cdot \vec{S}_j^\perp - 2 S_i^z S_j^z \pm \sqrt{3} {\,} \vec{z}  \cdot (\vec{S}_i \times \vec{S}_j)\right]
\label{eq:X2}
\end{align}
where $J=J_{z}=-2J_{\perp}=\mp 2D/\sqrt{3}>0$. Since $\mathcal{H}_{\rm HAF}$ and $\mathcal{H}_{\mathcal{X}^{\pm}}$ have the same form (see Eqs.~(\ref{eq:HAF}) and~(\ref{eq:X1})), spin configurations connected by the one-to-one mappings $\Gamma^{\mathbf{B}_i^{\pm}}$ necessarily have the same energy in their respective Hamiltonians. Hence, the HAF, $\mathcal{X}^{-}$ and $\mathcal{X}^{+}$ models have the same energy spectrum and thus the same extensive ground-state degeneracy. But the spin rotation confer them very peculiar signatures when probed magnetically.

The ground-state manifold of the Heisenberg antiferromagnet supports algebraic spin correlations~\cite{Garanin99a}. In neutron scattering measurements, these correlations take the form of anisotropic diffuse scattering known as ``pinch points'' (see Ref.~[\onlinecite{Henley10a}] for a pedagogical review by Chris Henley). As depicted in Fig.~\ref{fig:Sq}, pinch-point singularities are clearly visible in the structure factors of the $\mathcal{X}^{\pm}$ ground-state manifolds. The striking similarity of the HAF and $\mathcal{X}^{\pm}$ structure factors is actually a quantitative illustration of their underlying equivalence. But because the planar spin components are respectively antiferromagnetically and ferromagnetically coupled in $\mathcal{H}_{\rm HAF}$ and $\mathcal{H}_{\mathcal{X}^{\pm}}$, their collective fluctuations induce reversed spin correlations. This provides a noticeable example of pinch points induced by continuous ferromagnetic fluctuations.

As $T\rightarrow 0^{+}$, the $\mathcal{X}^{\pm}$ models are expected to undergo the same thermal order-by-disorder selection as the Heisenberg antiferromagnet~\cite{Chalker92a}, with the additional flavour that the octupolar order~\cite{Zhitomirsky08a} now bears a finite vector chirality.\\


\section{A connected family of spin liquids}

Spin chirality takes multiple forms. The non-collinearity of spins is directly measured by the vector chirality $\vec \chi_{ij}=\vec S_{i} \times \vec S_{j}$. For triangular units, one can further define a scalar chirality $\chi_{ijk}=\vec S_{i} \cdot\left(\vec S_{j} \times \vec S_{k}\right)$ which is a measure of the solid angle formed by the three spins. Vector chirality comes from the spin current involved in the strong magneto-electric coupling of some multiferroics~\cite{Katsura05a} and the emergence of skyrmion excitations~\cite{Muhlbauer09a}. As for scalar chirality, it can induce anomalous Hall effect when coupled to itinerant electrons~\cite{Ohgushi00a,Yoshii00a,Taguchi01a,Machida10a}.

While vector-chirality is intrinsically induced by the Dzyaloshinskii-Moriya term, we do not expect any long-range scalar-chiral order in the $\mathcal{X}^{\pm}$ models since the HAF spin liquid does not break $\mathcal{T}$ symmetry at finite temperature. It is thus tantalizing to see if, by taking advantage of the present three-fold mapping, it were possible to tune the Hamiltonians and induce scalar chirality spontaneously.

\begin{figure}[t!]
\begin{center}
\includegraphics[width=\columnwidth]{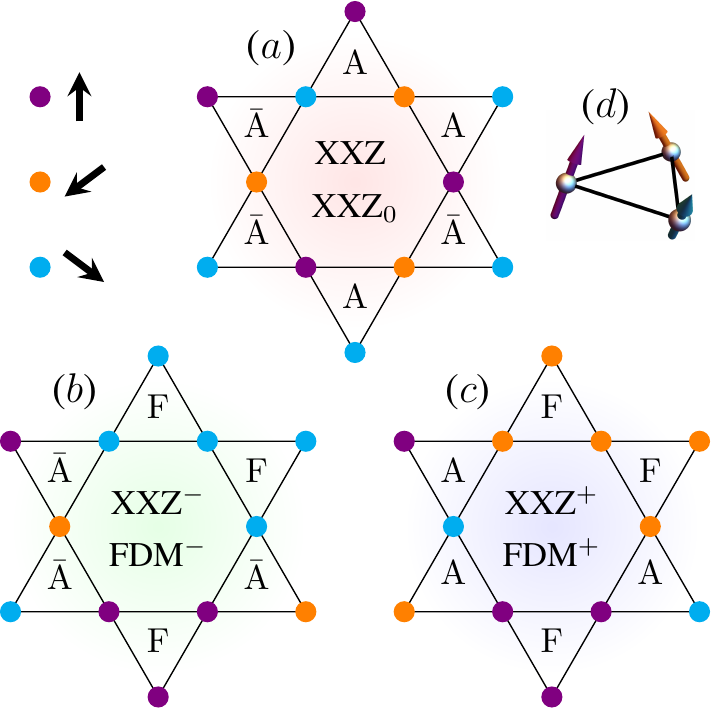}
\end{center}
\caption{
\textbf{Three-coloring ground states} -- The classical ground-state ensembles of the XXZ model and its chiral counterparts XXZ$^{\pm}$ are equivalent to the three-coloring problem, up to a global O(2) symmetry. ($a$) The equivalence is transparent for the XXZ ground state~\cite{Huse92a}, where each triangle possesses the three possible spin orientations rotated by $2\pi/3$ from each other, and $S^{z}=0$. The color code of the spin orientations is given in the top-left legend. The two antiferromagnetic permutations A$=\{\textcolor{violet}{\bullet},\textcolor{orange}{\bullet},\textcolor{cyan}{\bullet}\}$ and $\bar{\rm A}=\{\textcolor{violet}{\bullet},\textcolor{cyan}{\bullet},\textcolor{orange}{\bullet}\}$ are possible, giving a zero vector-chirality on average. ($b-c$) In this context, the apparition of vector-chirality in the XXZ$^{\pm}$ ground states is understood as the suppression of either the A or $\bar{\rm A}$ configurations in favor of a collinear state (F). The same scenario holds for the XXZ$_{0}$ and FDM$^{\pm}$ ground states where the finite out-of-plane magnetization makes the chirality scalar. An example of spin configuration with finite scalar chirality is given in panel ($d$): the planar projection of the spins corresponds to configuration A.
}
\label{fig:3color}
\end{figure}

Since our three-fold mapping does not affect the $z-$axis (see Fig.~\ref{fig:Map}), decreasing $J_{z}$ has the same influence on the HAF and $\mathcal{X}^{\pm}$ Hamiltonians: it tunes the HAF into the XXZ model
\begin{align}
\mathcal{H}_{\rm XXZ} = J \sum\limits_{\langle i,j\rangle}\left[ \vec{S}_i^\perp \cdot \vec{S}_j^\perp + \delta\, S_i^z S_j^z \right]
\label{eq:XXZ}
\end{align}
which is mapped onto what we name the XXZ$^{\pm}$ models
\begin{align}
\mathcal{H}_{\textrm{XXZ}^{\pm}}
&= -\frac{J}{2} \sum_{\langle i,j\rangle}\left[ \vec{S}_i^\perp \cdot \vec{S}_j^\perp - 2\delta\, S_i^z S_j^z \pm \sqrt{3} {\,} \vec{z}  \cdot (\vec{S}_i \times \vec{S}_j)\right]
\label{eq:XXZpm}
\end{align}
with $J>0$. This mapping is valid for all $\delta$ but for $-1/2 < \delta < 1$, the XXZ ground state remains a sub-ensemble of the HAF one ($\delta =1$) where all spins lie in plane ($\chi_{ijk}=0$). This model is equivalent to the three-coloring problem up to a global O(2) symmetry~\cite{Huse92a}, whose degeneracy is countable and extensive~\cite{Baxter70a}. As illustrated in Fig.~\ref{fig:3color}, the system is entirely paved with only two kinds of triangular configurations, A and $\bar{\rm A}$, with opposite vector-chirality~\cite{Huse92a}. The noticeable consequence of our mapping is that in the XXZ$^{-}$ and XXZ$^{+}$ ground states, the A and $\bar{\rm A}$ configurations are respectively replaced by a collinear state F with zero chirality; the resulting imbalance ensures finite and opposite vector-chirality between the two XXZ$^{\pm}$ ground states, while preserving their extensive degeneracy. From this point of view, it is interesting to think of the XXZ ground state as coming from the cancelation of positive and negative Dzyaloshinskii-Moriya terms, once ferromagnetism has been taken out.\\


\section{Chiral spin liquids}

On the other hand, for $\delta = -1/2$ Dzyaloshinskii-Moriya interactions become perfectly balanced by \textit{isotropic ferromagnetic} coupling
\begin{align}
\mathcal{H}_{\textrm{FDM}^{\pm}}
&= -\frac{J}{2} \sum_{\langle i,j\rangle} \left[\vec{S}_i \cdot \vec{S}_j \pm \sqrt{3} {\,} \vec{z}  \cdot (\vec{S}_i \times \vec{S}_j)\right].
\label{eq:FDM}
\end{align}
We name them the FDM$^{\pm}$ models. As a consequence, for each triangle, both the Dzyaloshinskii-Moriya (DM) induced~\cite{Elhajal02a} and ferromagnetic (FM) ground-state configurations minimize the classical energy
\begin{align}
&\quad\vec S_{\ell=\{0,1,2\}}=(\sin\theta\cos\phi_{\ell}^{\pm},\sin\theta\sin\phi_{\ell}^{\pm},\cos\theta),
\label{eq:FDMconfig}\\
&\left\{
\begin{matrix}
DM: \phi_{\ell}^{\pm}=\phi&\pm\frac{2\pi}{3}\ell& \Rightarrow\chi_{012}=&\pm&\frac{3\sqrt{3}}{2}\cos\theta\sin\theta^2\\
FM: \phi_{\ell}^{\pm}=\phi& &\Rightarrow\chi_{012}=&0&
\end{matrix}
\right.\nonumber
\end{align}
where the $\pm$ index distinguishes the two FDM$^{\pm}$ models. With respect to the XXZ$^{\pm}$ models where $\theta$ was imposed to be zero, the global degeneracy of the FDM$^{\pm}$ ground states is enhanced to O(3). Thus, while the $\vec S^{\perp}$ degrees-of-freedom conserves the character of a classical spin liquid, with the extensive degeneracy and algebraic correlations of the three-coloring problem, $S^{z}=\cos\theta$ can now take a finite uniform value, conferring a finite scalar chirality to any ground-state configuration with $\theta\neq\{0,\pi/2,\pi\}$ (see Eq.~(\ref{eq:FDMconfig})).

The three-fold mapping transforms the FDM$^{\pm}$ models back into the XXZ$_{0}$ Hamiltonian of Eq.~(\ref{eq:XXZ}) with $\delta = -1/2$, where the scalar chirality has vanished but the enhanced global O(3) degeneracy remains. It is noteworthy that the end point value $\delta = -1/2$ takes an elegant meaning along the XXZ$^{\pm}$ lines, namely that the ferromagnetic coupling becomes isotropic, which is hidden if only considering the XXZ model.\\

The emergence of scalar chirality in what is essentially a ``simple'' ferromagnet with Dzyaloshinskii-Moriya interactions is quite remarkable, with a rich potential for unconventional phenomena. For example the interplay between a chiral spin liquid and itinerant electrons is an up-and-coming topic~\cite{Udagawa12a,Chern13a,Gong14a}. Indeed, the FDM$^{\pm}$ ground state is neither fully ordered like a solid, or paramagnetic like a gas. In a pictorial way it is a magnetic liquid where strong correlations and fluctuations co-exist, which can then couple via double-exchange to another ``fluid'' made of itinerant electrons. While hopping on the scalar-chiral spin texture, the itinerant electrons pick up a Berry phase that might not only induce anomalous Hall conductivity~\cite{Ohgushi00a,Tatara02a,Taillefumier06a,Gong14a}, but at the same time feed back into the strongly correlated spin texture to induce exotic magnetic order~\cite{Martin08a,Akagi10a,Gong14a,Ishizuka15a}. This feedback actually does not require scalar chirality and would also be pertinent to the XXZ$^{\pm}$ lines.

It should be noted that given the large value of $D=\sqrt{3}J$, an experimental realization of the FDM$^{\pm}$ models \textit{per se} would arguably be difficult in solid state physics, but on the other hand, particularly promising for optical lattices. Indeed, the kagome geometry~\cite{Jo12a} and spin anisotropy~\cite{Struck13a} have been experimentally realized with ultracold atoms. There is also good hope that the active research on synthetic gauge fields might be able to produce synthetic Dzyaloshinskii-Moriya interactions~\cite{Cai12a,Cole12a,Radic12a}, with the caveat that the Dzyaloshinskii-Moriya vector should be out-of-plane here.

Last but not least, ferromagnetic insulators with Dzyaloshinskii-Moriya coupling have been studied in the context of magnon Hall effect, \textit{i.e.} where a transverse heat current is induced by a temperature gradient. It is intriguing to notice that the FDM$^{\pm}$ sits at the frontier between two different topological phases, indicating the closing of a gap between two magnon bands~\cite{Mook14a,Mook14b}. In light of our present work, and since the topological phase for $D<\sqrt{3}J$ is the same down to $D=0$~\cite{Mook14a,Mook14b}, it would be of great interest for future work to study the finite temperature physics of the Dzyaloshinskii-Moriya ferromagnet. This is especially true since chiral magnonic edge states and topological skyrmion excitations have been observed in simulations for $D/J\sim 0.4$~[\onlinecite{Pereiro14a}].\\

\begin{figure}[t!]
\centering
\includegraphics[width=\columnwidth]{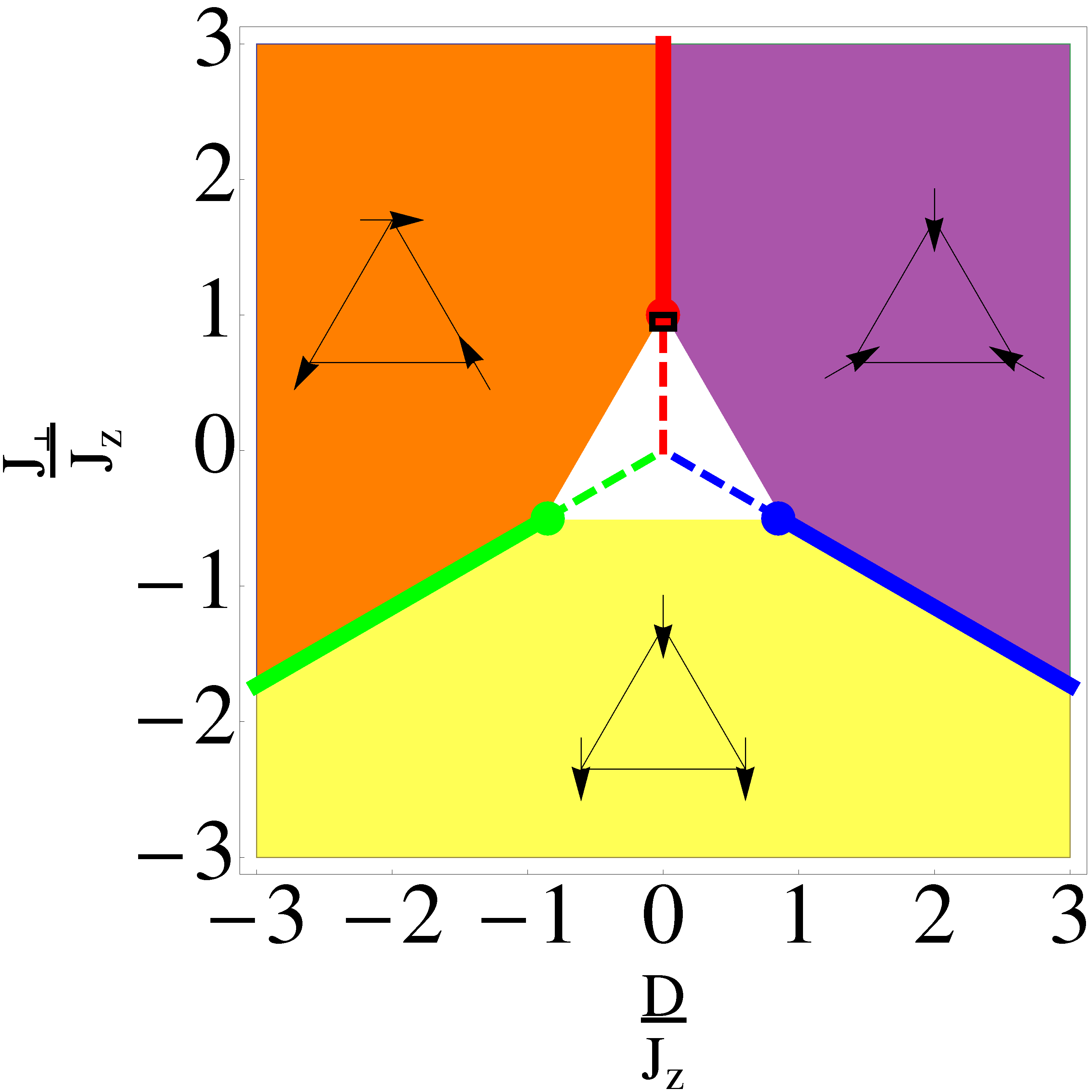}
\caption{\textbf{Phase diagram of Hamiltonian~(\ref{eq:ham}) obtained from linear spin wave theory for $J_{z}>0$} -- The HAF and $\mathcal{X}^{\pm}$ spin liquids are marked by dots and the yellow, orange and purple regions are long-range ordered phases. The white triangle delimits a regime where quantum corrections to the order parameters diverge, indicating a possible extended region of quantum disorder. In particular, DMRG has shown that the entire XXZ model (solid and dashed red lines) is a quantum spin liquid. Our three-fold transformation maps this quantum spin liquid onto the XXZ$^{\pm}$ models of Hamiltonian~(\ref{eq:XXZpm}) (green and blue lines) for $0<\delta<1$ (solid) and $\delta >1$ (dashed), which are thus also quantum spin liquids. Experimentally, independent parametrizations of the Herbertsmithite compound~\cite{Zorko08a,Han12a} put it at the tip of the white-triangle region (black rectangle).
}
\label{fig:PD}
\end{figure}
\section{Quantum fluctuations}

Our analysis has been so far focused on classical spins in order to precisely determine the nature of their classical ground states. However, it is important to keep in mind that our present three-fold mapping is also exact with quantum spins, since the local transformations are proper rotations, \textit{i.e.} \textit{unitary} matrices, and therefore preserve the commutation relations of the spin components.

We have investigated the consequences of quantum fluctuations for these spin liquids in the framework of linear spin wave theory within the parameters of Hamiltonian~(\ref{eq:ham}). These results are illustrated in Fig. \ref{fig:PD}. Approaching any of the HAF or $\mathcal{X}^{\pm}$ models (marked by dots), the linear spin wave Hamiltonian takes the same form, which simply confirms the equivalence of these three spin liquids in the presence of quantum fluctuations. The linear spin wave calculation also indicates the likelihood of quantum disorder around the center of Fig.~\ref{fig:PD}. Approaching the white-triangle region, a flat band of excitations collapses to zero energy leading to a divergence in the quantum correction to the order parameter.\\

Within the phase diagram of Fig.~\ref{fig:PD}, the $J_{\perp}=J_{z}=J>0$ line has drawn substantial interest for its relevance to Herbertsmithite ZnCu$_3$(OH)$_6$Cl$_2$ where Dzyaloshinskii-Moriya interactions are not negligible ($D/J\sim 0.044 - 0.08$)~\cite{Rigol07a,Rigol07c,Zorko08a,Shawish10a}. We reproduce the results of Refs.~[\onlinecite{Elhajal02a},\onlinecite{Ballou03a}] done on this line of parameters, namely that classically~\cite{Elhajal02a} and up to linear order in spin wave theory~\cite{Ballou03a}, magnetic order is stabilized for any finite value of $D$. However, higher order terms in quantum fluctuations studied by Exact Diagonalization~\cite{Cepas08a,Rousochatzakis09a}, Schwinger-boson~\cite{Messio10a,Huh10a} and perturbative methods~\cite{Tovar09a} have shown that quantum disorder actually persists over a finite region up to $D/J\sim 0.1$, which includes Herbertsmithite. Our goal here is not to claim explaining the spin liquid nature of Herbertsmithite which has been extensively studied, but rather to set our theory on an experimental footing. In particular, it should be noted that at linear order in quantum fluctuations, the small XXZ anisotropy observed in Herbertsmithite~\cite{Han12a,Rigol07c} ($J_{\perp}/J_{z}\approx 0.9$) brings this compound at the tip of the white-triangle region with quantum disorder.

Over the past year, the XXZ line with $D=0$ and $0<J_{z}<J_{\perp}$ has also received significant attention for the spin-liquid nature of its ground state for spins $S=1/2$~\cite{Gotze15a,He15a,Hu15a,Zhu15a}, and the complex quantum order-by-disorder mechanism that takes place for $S>1/2$~\cite{Chernyshev14a,Gotze15a}. Noticeably for spin$-1/2$, the density matrix renormalization group (DMRG) approach indicates that the quantum spin liquid persists for $0<J_{\perp}/J_{z}<1$ and $D=0$~\cite{He15a}. The consequences on our work are multiple. First of all, our three-fold mapping makes the XXZ$^{\pm}$ models \textit{quantum spin liquids} for spin$-1/2$ (see green/blue solid and dashed lines in Fig.~\ref{fig:PD} and $\delta>0$ in Eq.~(\ref{eq:XXZpm})). Furthermore, the central point of our phase diagram $D=J_{\perp}=0$ has Ising anisotropy and is known for remaining a quantum paramagnet even for arbitrarily small transverse fields~\cite{Moessner00b,Moessner01b,Nikolic05a,Powalski13a}. As such, it has been described as a rare example of ``disorder-by-disorder'', a mechanism proposed by Fazekas and Anderson~\cite{Fazekas74a} where quantum fluctuations select a disordered sub--manifold of the classically degenerate ground state. Within the framework of our three-fold mapping, this remarkable resistance to order can be understood as the consequence of being at the intersection of three (dashed) lines of spin liquids, in a way reminiscent of what has been observed in pyrochlore systems~\cite{Yan13a}. However, please note that the spin$-1/2$ phase diagram is expected to be highly anisotropic around this central point, since it has been shown to order into a superfluid phase for $D=0$ and $J_{\perp}<0$~\cite{Isakov06a,He15a}, and thus also along the symmetric lines $J_{\perp}=\pm D/\sqrt{3}>0$ around the origin according to our three-fold mapping.\\

\section{Conclusion}

We have discovered a connected network of quantum spin liquids on the kagome lattice, which are mapped onto each other via a three-fold transformation. One of the branches of this network is the anisotropic XXZ model, known to be a quantum spin liquid for spin$-1/2$~\cite{Gotze15a,He15a,Hu15a,Zhu15a} (see the red lines in Figs.~\ref{fig:Map} and~\ref{fig:PD}), which includes the actively studied Heisenberg antiferromagnet. While every triad of Hamiltonians connected by this mapping have exactly the same energy spectrum at the classical and quantum level, their spin configurations are transformed. As a consequence, the three-fold mapping of the XXZ line gives rise to new spin liquids with intrinsic vector chirality because of Dzyaloshinskii-Moriya interactions. The Ising antiferromagnet sits at the centre of this map (see Fig.~\ref{fig:PD}), which sheds a new light on the unique propensity of this model to remain disordered~\cite{Moessner00b,Moessner01b,Nikolic05a,Powalski13a}.

Beyond these three branches of quantum spin liquids, we have studied the stability of Hamiltonian~(\ref{eq:ham}) for $J_{z}>0$, up to linear order in spin wave theory. We have found an extended region of the phase diagram in Fig.~\ref{fig:PD} where quantum disorder prevails. The small XXZ anisotropy observed in Herbertsmithite~\cite{Han12a,Rigol07c} ($J_{\perp}/J_{z}\approx 0.9$) brings this compound within the tip of this extended region.

At the classical level, the Heisenberg antiferromagnet maps onto two models where algebraic correlations take the form of ferromagnetic pinch points visible in the structure factor of Fig.~\ref{fig:Sq}. Keeping the Dzyaloshinskii-Moriya term constant, if one tunes the $J_{z}$ coupling of these models until they become isotropic ferromagnet, the chirality can spontaneously become scalar.\\


Our work opens a wide range of exciting directions to follow, both by theorists and experimentalists. In light of the intense research on the Heisenberg antiferromagnet and XXZ models, here we propose two lines of systems with the same energy spectra, but different (chiral) magnetic signatures. With this new probe at hand and Figs.~\ref{fig:Map} and~\ref{fig:PD} in mind, it would be of great interest to look for new insights as one approaches these models and their chiral counterparts from different angles in parameter space ($J_{\perp},J_{z},D$). In particular, the spreading of quantum disorder within the white triangle of Fig.~\ref{fig:PD} and in its vicinity shall conserve the three-fold symmetry, and be mediated by quantum order-by-disorder mechanisms as we vary the spin length $S$~\cite{Chernyshev14a,Gotze15a}.

The inclusion of 2$^{\rm nd}$ and 3$^{\rm rd}$ nearest-neighbour interactions $J_{2}=J_{3}=J_{NNN}$ is known to stabilize a chiral spin liquid at finite value $J_{NNN}^{c}$~\cite{Messio12a,Gong14a,He14a,Gong15a,He15a,Zhu15a}. This value $J_{NNN}^{c}$ has been shown by DMRG to decrease as the antiferromagnetic $J_{z}$ coupling vanishes. This means that the chiral spin liquid is getting closer to the nearest-neighbour XXZ model as $J_{z}$ goes from 1 to 0. It would thus be very tempting to extend this work to ferromagnetic coupling ($J_{z}<0$) towards the XXZ$_{0}$ and equivalent FDM$^{\pm}$ points. Since $\mathcal{T}$ symmetry can be spontaneously broken in the classical FDM$^{\pm}$ ground states, the possible connection with the chiral spin liquids at finite $J_{NNN}$ is a particularly attractive open question.

Beyond kagome physics, the present methodology can be applied to a broad range of lattices and dimensions. Our results especially suggests that systems supporting the ``disorder-by-disorder'' mechanism~\cite{Fazekas74a}, such as the Ising antiferromagnet here~\cite{Moessner00b}, are good places to look for hidden spin liquids in the neighbouring parameter space.\\

On the experimental front, our work fits within the on-going effort for the experimental realization of frustrated systems in optical lattices~\cite{Jo12a,Struck13a}, and especially to produce tunable synthetic Dzyaloshinskii-Moriya interactions~\cite{Cai12a,Cole12a,Radic12a}.

We hope that our results will further motivate experimental efforts on the synthesis and characterization of kagome materials with anisotropic nearest-neighbour interactions. The recently synthesized ternary intermetallic compounds Dy$_{3}$Ru$_{4}$Al$_{12}$~\cite{Gorbunov14a} and Yb$_{3}$Ru$_{4}$Al$_{12}$~\cite{Nakamura15a} are very promising materials to start with, since the 4f orbitals of rare-earth ions are known to induce very anisotropic and short-range interactions. Furthermore the presence of itinerant electrons make them natural materials to probe the chirality of the underlying spin texture. Their crystal structure, however, corresponds to a distorted kagome lattice. To impose kagome symmetry is a chemistry challenge, but such was the case for Volborthite Cu$_{3}$V$_{2}$O$_{7}$(OH)$_{2}$ $\bullet$ 2H$_{2}$O, 14 years ago~\cite{Hiroi01a}, which predated the synthesis of a growing number of materials with essentially perfect kagome symmetry~\cite{Shores05a,Colman08a,Okamoto09a,Aidoudi11a,Han14a}. According to our three-fold mapping, the places to look for would be large antiferromagnetic $J_{z}$, as well as around the XXZ$_{0}$ point where only small Dzyaloshinskii-Moriya terms are required. In light of Refs.~[\onlinecite{Pereiro14a},\onlinecite{Mook14a}], the region neighbouring the FDM$^{\pm}$ models is also very promising, even for smaller values of $D$ and anisotropic $J_{z}$. At the proximity of these high-symmetry points, especially the one at the centre of the white triangle, chemical, hydrostatic and uni-axial pressure might help the exploration of the phase diagram, as done in rare-earth pyrochlore oxides~\cite{Wiebe15a}.

\begin{acknowledgments}
It is a pleasure to thank Nic Shannon, Mathieu Taillefumier, Andreas Laeuchli and Andriy Nevidomskyy for useful discussions. This work was supported by the Okinawa Institute of Science and Technology Graduate University.
\end{acknowledgments}
\bibliography{/Users/Ludo/Desktop/biblio.bib}
\end{document}